\documentclass[12pt]{article}
\usepackage{graphicx}

\begin{document}

\baselineskip 0.1667in

\begin{center}
{\large \textbf{Reexamination of Barnett's Experiment Based on}}

{\large \textbf{the Modified Lorentz Force Law}}

\vspace{1cm}

\textsf{Ching-Chuan Su}

Department of Electrical Engineering

National Tsinghua University

Hsinchu, Taiwan

\vspace{1cm}
\end{center}

\noindent \textbf{Abstract}\ -- Barnett's experiment demonstrates that the
induction on a stationary cylindrical capacitor in the presence of a
rotating magnet or solenoid is zero. In this investigation, based on the
modified Lorentz force law, which complies with Galilean transformations and
depends on relative velocities, the induction on the capacitor is
reexamined. When the rotating solenoid is long and the capacitor is placed
inside the solenoid, it is seen that the induction actually vanishes as
observed in Barnett's experiment. However, when the capacitor is placed
outside the solenoid or when the solenoid is short, it is shown that the
induction can departure from zero. This prediction provides a means to test
the validity of the modified Lorentz force law.

\vspace{1.5cm}

\noindent {\large \textbf{1. Introduction}}\\[0.2cm]
It is known that in as early as 1831 Faraday demonstrated the unipolar
induction on a metallic wire which is rotating in the presence of a magnet,
where the rotation axis is parallel to the magnetic field. Meanwhile, it
seems that the induction is determined by the relative motion between the
wire and the magnet. Thus it is expected that a similar induction can be
observed for a stationary wire and a rotating magnet or solenoid carrying a
current. In 1912 Barnett conducted experiments to justify this issue of
relative motion [1]. In Barnett's experiment two coaxial conducting
cylinders of different radii are placed inside a solenoid of larger radius
concentrically (see Fig.\hspace{0.1cm}1). The two cylinders which form a
capacitor are connected to an electrometer with conducting wires. The
solenoid was made to rotate about its center axis and to carry a current.
The induced \textrm{electromotive force} will drive charges of opposite
signs on to the two cylinders of the capacitor via the connecting wires,
respectively. By using switches which provide suitable electric insulation,
the charge on the capacitor can be measured and then the induction can be
determined. However, for the case where the solenoid is rotating and the
capacitor together with the connecting wires is stationary, the observed
charge is merely a minute fraction of the expected amount, within the limits
of experimental errors [1]. Without using the switches, Kennard also tried
to examine the effect of relative motion by measuring the potential
difference across a similar cylindrical capacitor placed inside or outside
the concentric solenoid [2, 3]. Negative results were also reported, except
for the disturbances which were attributed to the induction when the
magnetization was reversed [3]. Thus it seems that the principle of
relativity which states that physical laws depend on relative motion is
violated. On the other hand, it is argued that the principle of relativity
applies for the relative motion of translation, but not for the relative
motion of rotation. Discussions on the effects of relative rotational motion
on the induction still remain active in the literature [4-6]. But this issue
seems not yet solved conclusively with quantitative analysis.

In this investigation we present a reexamination of Barnett's experiment by
resorting to the modified Lorentz force law, which is derived from a wave
equation in a quantum mechanical way based on the local-ether model of wave
propagation [7]. This local-ether wave equation in turn leads to a unified
quantum theory of the gravitational and electromagnetic forces in
conjunction with the origin and identity of the gravitational and inertial
mass. Furthermore, this wave equation accounts for a wide variety of
phenomena in modern physics, including the Sagnac effect with wave
propagation, Fizeau's experiment with moving media, the gravitational
redshift in the Pound-Rebka experiment, and the Hafele-Keating experiment
with fast-moving atomic clocks [7]. For quasi-static cases, the modified
force law is not new, as it can be derived from the Riemann force law which
in turn has been proposed in as early as 1861 and can reduce to the Lorentz
force law under some common situations [8].

As well as the Riemann force, the modified Lorentz force law is in
compliance with Galilean transformations and can be in accord with the
principle of relativity as it depends on relative velocities. Qualitatively,
these force laws immediately account for the difference between the relative
motions of rotation and translation. For two small objects, a relative
motion of rotation is equivalent to a relative motion of translation at a
given instant. However, if either object is large, the situation can be
different. When the aforementioned solenoid is rotating and the connecting
wires along with the capacitor are stationary, the relative motions between
a given segment of the wires and different segments of the solenoid are
different in direction. This is different from the case where the capacitor
is rotating and the solenoid is stationary, as the relative motions between
a given rotating segment and various segments of the solenoid then become
identical at a given instant. Based on the modified Lorentz force law, the
vanishing induction with a rotating solenoid can be accounted for without
the breakdown of the principle of relativity. Further, a fundamentally
different consequence is pointed out. That is, the induction on the
stationary capacitor can be different from zero when the capacitor is placed
outside the rotating solenoid or when the solenoid is short in length. This
prediction provides a means to test the validity of the Riemann force law
and the modified law.

\vspace{1cm}

\noindent {\large \textbf{2. Modified Lorentz Force Law}}\\[0.2cm]
It is proposed that the electromagnetic force exerted on a charged particle
due to various source particles can be given in terms of the augmented
scalar potential. Precisely, the augmented scalar potential $\breve{\Phi}$
due to source particles of charge density $\rho _{v}$ and experienced by the
effector particle is given explicitly by the integral over a volume
containing all the source particles 
$$
\breve{\Phi}(\mathbf{r},t)=\frac{1}{4\pi \epsilon _{0}}\int \left( 1+\frac{%
v_{es}^{2}}{2c^{2}}\right) \frac{\rho _{v}(\mathbf{r}^{\prime },t-R/c)}{R}%
dv^{\prime },\eqno
(1) 
$$
where $v_{es}=|\mathbf{v}_{es}|$, the velocity difference $\mathbf{v}_{es}=%
\mathbf{v}_{e}-\mathbf{v}_{s}$, $\mathbf{v}_{e}$ is the velocity of the
effector located at position $\mathbf{r}$ at instant $t$, $\mathbf{v}_{s}$
is that of the source particles distributed at position $\mathbf{r}^{\prime
} $ at an earlier instant $t^{\prime }$ ($=t-R/c$), the propagation range $R$
($=|\mathbf{r}-\mathbf{r}^{\prime }|$) is the distance from the source point 
$\mathbf{r}^{\prime }$ to the field point $\mathbf{r}$, and $R/c$ denotes
the propagation delay time from the source to the effector at the respective
positions and instants.

Further, it is postulated that the electromagnetic force exerted on the
effector of charge $q$ is given in terms of the augmented scalar potential
by [7] 
$$
\mathbf{F}(\mathbf{r},t)=q\left\{ -\nabla \breve{\Phi}(\mathbf{r},t)+\left( 
\frac{\partial }{\partial t}\sum\limits_{i}\hat{\imath}\frac{\partial }{%
\partial v_{ei}}\breve{\Phi}(\mathbf{r},t)\right) _{e}\right\} ,\eqno
(2) 
$$
where $v_{ei}=\mathbf{v}_{e}\cdot \hat{\imath}$, $\hat{\imath}$ is a unit
vector, the index $i=x,y,z$, and the time derivative $(\partial $$/\partial
t)_{e}$ is referred to the effector frame with respect to which the effector
is stationary. Physically, an effector-frame time derivative represents the
time rate of change in some quantity experienced by the effector. It is
noticed that this formula resembles Lagrange's equations adopted by \textrm{%
Weber}, Riemann, and Thomson in the early development of electromagnetic
force [9, 10]. And the augmented scalar potential resembles the
velocity-dependent potential energy introduced by \textrm{Riemann} [9].

Ordinarily, the magnetic force is due to a conduction current where the
mobile charged particles forming the current are actually embedded in a
matrix, such as electrons in a metal wire. The ions that constitute the
matrix tend to electrically neutralize the mobile particles and thus the
conduction current is neutralized. Furthermore, the mobile source particles
drift very slowly with respect to the matrix. Thereby, as shown in [7], the
proposed electromagnetic force law (2) based on the augmented scalar
potential can then be given in terms of the electric scalar potential $\Phi $
and the magnetic vector potential $\mathbf{A}$ by 
$$
\mathbf{F}(\mathbf{r},t)=-q\nabla \Phi (\mathbf{r},t)+q\nabla \left[ \mathbf{%
v}_{em}\cdot \mathbf{A}(\mathbf{r},t)\right] -q\left( \frac{\partial }{%
\partial t}\mathbf{A}(\mathbf{r},t)\right) _{e},\eqno
(3) 
$$
where $\mathbf{v}_{em}$ is the velocity of the effector particle referred
specifically to the matrix which in turn is supposed to move as a whole at a 
\textrm{uniform} velocity $\mathbf{v}_{m}$. The scalar and the vector
potential in turn are given explicitly in terms of the net charge density $%
\rho _{n}$ and the neutralized current density $\mathbf{J}_{n}$ respectively
by the volume integrals 
$$
\Phi (\mathbf{r},t)=\frac{1}{4\pi \epsilon _{0}}\int \frac{\rho _{n}(\mathbf{%
r}^{\prime },t)}{R}dv^{\prime }\eqno
(4) 
$$
and 
$$
\mathbf{A}(\mathbf{r},t)=\frac{\mu _{0}}{4\pi }\int \frac{\mathbf{J}_{n}(%
\mathbf{r}^{\prime },t)}{R}dv^{\prime },\eqno
(5) 
$$
where $\mathbf{J}_{n}=\rho _{v}\mathbf{v}_{sm}$, $\mathbf{v}_{sm}$ is the
velocity of the mobile source particles referred specifically to the matrix,
and the propagation time is neglected as the cases considered in this
investigation are quasi-static.

By using Galilean transformations the preceding force law can be given by 
$$
\mathbf{F}(\mathbf{r},t)=-q\nabla \Phi (\mathbf{r},t)+q\nabla \left[ \mathbf{%
v}_{em}\cdot \mathbf{A}(\mathbf{r},t)\right] -q\left( \frac{\partial }{%
\partial t}\mathbf{A}(\mathbf{r},t)\right) _{m}-q\left( \mathbf{v}_{em}\cdot
\nabla \right) \mathbf{A}(\mathbf{r},t),\eqno
(6) 
$$
where the time derivative $(\partial $$\mathbf{A}/\partial t)_{m}$ is
referred to the matrix frame with respect to which the matrix is stationary.
The terms $-q\nabla \Phi $ and $-q(\partial \mathbf{A}/\partial t)_{m}$
represent the electrostatic force and the electric induction force,
respectively; meanwhile, the terms $q\nabla (\mathbf{v}_{em}\mathbf{\cdot A}%
) $ and $-q(\mathbf{v}_{em}\mathbf{\cdot }\nabla )\mathbf{A}$ represent the
magnetostatic force and the magnetic induction force, respectively. By using
a vector identity, the two force terms associated with the effector velocity 
$\mathbf{v}_{em}$ can be combined into the magnetic force. Thus the
electromagnetic force exerted on an effector particle becomes a more
familiar form 
$$
\mathbf{F}(\mathbf{r},t)=q\left\{ -\nabla \Phi (\mathbf{r},t)-\left( \frac{%
\partial }{\partial t}\mathbf{A}(\mathbf{r},t)\right) _{m}+\mathbf{v}%
_{em}\times \nabla \times \mathbf{A}(\mathbf{r},t)\right\} .\eqno
(7) 
$$
According to the dependences of the force terms on $\mathbf{v}_{em}$, one is
led to express the force law in terms of the fields $\mathbf{E}$ and $%
\mathbf{B}$ in the form 
$$
\mathbf{F}(\mathbf{r},t)=q\left\{ \mathbf{E}(\mathbf{r},t)+\mathbf{v}%
_{em}\times \mathbf{B}(\mathbf{r},t)\right\} ,\eqno
(8) 
$$
where the electric field $\mathbf{E}$ and the magnetic field $\mathbf{B}$
are then defined explicitly in terms of $\Phi $ and $\mathbf{A}$ as 
$$
\mathbf{E}(\mathbf{r},t)=-\nabla \Phi (\mathbf{r},t)-\left( \frac{\partial }{%
\partial t}\mathbf{A}(\mathbf{r},t)\right) _{m}\eqno
(9) 
$$
and 
$$
\mathbf{B}(\mathbf{r},t)=\nabla \times \mathbf{A}(\mathbf{r},t).\eqno
(10) 
$$
The electromagnetic force law under the ordinary low-speed condition
represents modifications of the Lorentz force law, which comply with
Galilean transformations and can be in accord with the principle of
relativity as the involved velocities are relative.

The fundamental modifications are that the current density generating the
potential $\mathbf{A}$, the time derivative of $\mathbf{A}$ in the electric
induction force, and the effector velocity connecting to $\nabla \times 
\mathbf{A}$ in the magnetic force are all referred specifically to the
matrix frame. It is pointed out that this particular frame has been adopted
tacitly in common practice dealing with the magnetic force, such as with the
magnetic deflection. Further, the divergence and the curl relations for the
corresponding electric and magnetic fields are derived. Apart from some
minute deviation terms, these four relationships are just Maxwell's
equations, with the exception that the velocity determining the involved
current density and the associated time derivatives are also referred to the
matrix frame [7].

\vspace{1cm}

\noindent {\large \textbf{3. Electromotance}}\\[0.2cm]
We then go on to consider the electromotance (or called electromotive force)
induced on a conducting wire in the presence of a solenoid. For a
neutralized solenoid carrying a static current, both the electrostatic force
and the electric induction force vanish. Thus the force contributes to the
electromotance is the magnetic force 
$$
\mathbf{F}=q\mathbf{v}_{em}\times \mathbf{B.}\eqno
(11) 
$$
For the case where the solenoid is stationary and the wire is rotating at a
rate $\omega $ with respect to the center $z$ axis of the solenoid, the
effector velocity associated with the magnetic force is $\mathbf{v}_{em}=%
\hat{z}\omega \times \mathbf{r}$, where $\mathbf{r}$ is the directed radial
distance of the effector from the axis. Inside a long solenoid, the magnetic
field $\mathbf{B}$ is known to be uniform. Thus the electromotance $\mathcal{%
V}$ induced on the wire $C$ is given by 
$$
\mathcal{V}=\int_{C}\mathbf{v}_{em}\times \mathbf{B}\cdot d\mathbf{l}=\left. 
\frac{1}{2}\omega B_{0}r^{2}\right| _{r=r_{a}}^{r_{b}},\eqno
(12) 
$$
where $\mathbf{B}=\hat{z}B_{0}$ with $B_{0}$ being a constant and $r_{a}$
and $r_{b}$ are the radial distances of the endpoints of the wire from the
axis. Thus the electromotance is given by $\mathcal{V}=\omega
(r_{b}^{2}-r_{a}^{2})B_{0}/2$.

For the case where the solenoid is rotating and the conducting wire is
stationary, the velocity $\mathbf{v}_{em}$ varies among the various segments
around the rotating solenoid, as the matrix velocity is no longer uniform.
Consequently, the force law in terms of the vector potential $\mathbf{A}$ or
the magnetic field $\mathbf{B}$ is not applicable. Instead, the general form
of the magnetic force 
$$
\mathbf{F}=\frac{q\mu _{0}}{4\pi }\left\{ \nabla \left( \int \frac{\mathbf{v}%
_{em}\cdot \mathbf{J}_{n}}{R}dv^{\prime }\right) -\int (\mathbf{v}_{em}\cdot
\nabla )\frac{\mathbf{J}_{n}}{R}dv^{\prime }\right\} \eqno
(13) 
$$
or 
$$
\mathbf{F}=\frac{q\mu _{0}}{4\pi }\int \mathbf{v}_{em}\times \nabla \times 
\frac{\mathbf{J}_{n}}{R}dv^{\prime },\eqno
(14) 
$$
should be used, where the del operator applies only on the position vector $%
\mathbf{r}$ incorporated in $R$. For a thin wire $C^{\prime }$ carrying a
conduction current $I$, the current density is in the direction tangent to
the wire. Thus the electromotance is given by the double integral 
$$
\mathcal{V}=\frac{\mu _{0}I}{4\pi }\int_{C}\left\{ \int_{C^{\prime }}\nabla
\left( \frac{1}{R}\right) \mathbf{v}_{em}\cdot d\mathbf{l}^{\prime
}-\int_{C^{\prime }}\left( \mathbf{v}_{em}\cdot \nabla \right) \frac{1}{R}d%
\mathbf{l}^{\prime }\right\} \cdot d\mathbf{l,}\eqno
(15) 
$$
where the current is supposed to be static and uniform and is given by $%
I=\rho _{l}v_{sm}$, $\rho _{l}$ is the line charge density of the mobile
source particles (excluding the one of the neutralizing matrix), and $d%
\mathbf{l}^{\prime }$ is parallel to $\mathbf{v}_{sm}$. The preceding
formula may be difficult in calculation, as the matrix velocity and hence $%
\mathbf{v}_{em}$ are not fixed.

A simpler formula for electromotance can be obtained directly from the force
law (3). Consider the electromotance induced on a conducting wire $C$ due to
a linear short element of a conduction current and of directed length $%
\mathbf{l}^{\prime }$. Thus (3) leads to 
$$
\mathcal{V}=\frac{\mu _{0}I}{4\pi }\left\{ \int_{C}(\mathbf{v}_{em}\cdot 
\mathbf{l}^{\prime })\nabla \left( \frac{1}{R}\right) \cdot d\mathbf{l}%
-\int_{C}\frac{d}{dt}\left( \frac{\mathbf{l}^{\prime }}{R}\right) \cdot d%
\mathbf{l}\right\} ,\eqno
(16)
$$
where $R$ incorporated in the time derivative is varying with time whenever
there is a relative motion between the wire and the current element and thus
the time derivative is really associated with a quantity experienced by the
effector on the wire. It is noted that for a linear short wire of directed
length $\mathbf{l}$, 
$$
\frac{d}{dt}\left( \frac{\mathbf{l}^{\prime }}{R}\right) \cdot \mathbf{l}=%
\frac{d}{dt}\left( \frac{\mathbf{l}^{\prime }\cdot \mathbf{l}}{R}\right) -%
\frac{\mathbf{l}^{\prime }}{R}\cdot \frac{d\mathbf{l}}{dt}.\eqno
(17)
$$
The derivative $d\mathbf{l}/dt$ in turn corresponds to a rotation or
deformation and is equal to the difference of velocity between the endpoints
of the wire. The sum (designated as $K$) of the first integral in (16) and
the last term in (17) can be given by 
$$
K=\mathbf{\tilde{v}}\cdot \mathbf{l}^{\prime }\left( \frac{1}{R_{b}}-\frac{1%
}{R_{a}}\right) +\frac{1}{\tilde{R}}\mathbf{l}^{\prime }\cdot (\mathbf{v}%
_{b}-\mathbf{v}_{a}),\eqno
(18)
$$
where $\mathbf{v}_{a}$ and $\mathbf{v}_{b}$ denote the velocities of the
endpoints $a$ and $b$ of the wire with respect to the current element,
respectively, $R_{a}$ and $R_{b}$ are their distances from this element, and 
$\mathbf{\tilde{v}}$ or $\tilde{R}$ denotes a suitable mean value (say, the
arithmetic average) between the endpoints. Then it is easy to show that when
the directed length $\mathbf{l}$ approaches zero, one has 
$$
K=\frac{\mathbf{v}_{b}\cdot \mathbf{l}^{\prime }}{R_{b}}-\frac{\mathbf{v}%
_{a}\cdot \mathbf{l}^{\prime }}{R_{a}}.\eqno
(19)
$$
Thus the electromotance due to the current element can be given by 
$$
\mathcal{V}=\frac{\mu _{0}I}{4\pi }\left\{ \left. \frac{\mathbf{v}_{em}\cdot 
\mathbf{l}^{\prime }}{R}\right| _{a}^{b}\ -\ \frac{d}{dt}\left( \frac{%
\mathbf{l}^{\prime }\cdot \mathbf{l}}{R}\right) \right\} .\eqno
(20)
$$

Thereby, the electromotance induced over a wire $C$ due to a
current-carrying wire $C^{\prime }$, both of arbitrary shape and length, is
then given by 
$$
\mathcal{V}=\left. \frac{\mu _{0}I}{4\pi }\int_{C^{\prime }}\frac{\mathbf{v}%
_{em}\cdot d\mathbf{l}^{\prime }}{R}\right| _{a}^{b}\ -\ \frac{d}{dt}\int_{C}%
\mathbf{A}\cdot d\mathbf{l},\eqno
(21) 
$$
where $a$ and $b$ are the endpoints of $C$ and the vector potential $\mathbf{%
A}$ is 
$$
\mathbf{A}=\mu _{0}I\int_{C^{\prime }}\frac{1}{4\pi R}d\mathbf{l}^{\prime }%
\mathbf{.}\eqno
(22) 
$$
It is noted that the formula for the vector potential with a nonuniform
matrix velocity is still identical to the one with a uniform velocity, as
the matrix velocity does not affect the drift speed. Thus the electromotance
is composed of two terms, of which the one associated with $\mathbf{v}_{em}$
is primarily due to the magnetostatic force and the other is primarily due
to the induction force.

Like the work done by a conservative force, the electromotance due to the
magnetostatic force depends on the positions of the endpoints and not on the
shape of the path connecting them. For a closed wire $C$, this term tends to
vanish and thus the electromotance reduces to 
$$
\mathcal{V}=-\frac{d}{dt}\oint_{C}\mathbf{A}\cdot d\mathbf{l}=-\frac{d}{dt}%
\int_{S}\mathbf{B}\cdot d\mathbf{s},\eqno
(23) 
$$
where $S$ denotes the surface enclosed by the loop $C$, $d\mathbf{s}$ is in
the direction normal to the surface, and the magnetic field is still related
to the magnetic vector potential by (10), in spite of the matrix velocity
being nonuniform. This formula looks like the one which is commonly cited to
be related to Faraday's law of induction [11]. However, it is of essence to
note that owing to the incorporation of the term $d\mathbf{l}/dt$ in (17),
the time derivative in the preceding formula then operates on $d\mathbf{l}$
or $d\mathbf{s}$ and thus a rotation or deformation along with a translation
of the loop contributes to the electromotance, as well as the time variation
of $\mathbf{A}$ or $\mathbf{B}$ itself does.

Then consider the electromotance induced on a cylindrical capacitor in the
presence of a solenoid $C^{\prime }$ of circular shape and carrying a
uniform current $I$. Thus the resulting potential $\mathbf{A}$ is
azimuthally symmetric. When either the solenoid or the capacitor is or both
of them are rotating about the center axis of the solenoid, the integral of $%
\mathbf{A}$ in (21) is invariant with respect to time. Thereby, the
electromotance becomes an even simpler form 
$$
\mathcal{V}=\left. \frac{\mu _{0}I}{4\pi }\int_{C^{\prime }}\frac{1}{R}%
\mathbf{v}_{em}\cdot d\mathbf{l}^{\prime }\right| _{a}^{b},\eqno
(24) 
$$
where $a$ and $b$ denote two given points on the inner and outer cylinders,
respectively, as depicted in Fig.\hspace{0.1cm}1. The actual path $C$
implicitly associated with the preceding formula starts from point $a$ and
then traverses part of the inner cylinder, the connecting wires, and part of
the outer cylinder, and finally comes to point $b$. However, since this
electromotance does not differentiate the paths connecting the two
endpoints, the electromotance over the actual path is then identical to the
one over a complementary path connecting $a$ to $b$ directly via the space
between the two cylinders. For the case where the solenoid is rotating and
the capacitor is stationary, the relative motions between a given
constituent segment of the actual path and different segments of the
solenoid are different in direction. This situation is different from the
case where the capacitor is rotating and the solenoid is stationary.

For each segment of a closely wound solenoid which is rotating, $\mathbf{v}%
_{em}$ is along the azimuthal direction and $d\mathbf{l}^{\prime }$ is
almost along it. Further, the directions of these two vectors change in a
coordinated way around the solenoid and hence the product $\mathbf{v}_{em}%
\mathbf{\cdot }d\mathbf{l}^{\prime }$ in the preceding integral is a
constant. Thus, for the stationary capacitor and connecting wires, the
induced electromotance is given by 
$$
\mathcal{V}=-\frac{1}{2}\mu _{0}r_{0}\omega IN\left[
L(r_{b},z_{b})-L(r_{a},z_{a})\right] ,\eqno
(25) 
$$
where $\omega $ is the rotation rate, $N$ is the number of turns of the
solenoid, $r_{0}$ is its radius, $r_{a}$ and $z_{a}$ are the coordinates of
the endpoint $a$ in the cylindrical system, $r_{b}$ and $z_{b}$ are those of 
$b$, and $L$ denotes the contour integral of $1/R$ over the spiral structure
of the solenoid. Quantitatively, the dimensionless integral $L$ can be given
by 
$$
L(r,z)=\frac{r_{0}}{2\pi (z_{2}-z_{1})}\int_{z_{1}}^{z_{2}}\int_{0}^{2\pi }%
\frac{1}{\sqrt{(r_{0}\cos \phi ^{\prime }-r)^{2}+r_{0}^{2}\sin ^{2}\phi
^{\prime }+(z^{\prime }-z)^{2}}}d\phi ^{\prime }dz^{\prime },\eqno
(26) 
$$
where the solenoid is supposed to be located between $z_{1}$ and $z_{2}$. It
is seen that aside from a scaling factor the function $L$ is identical to
the electric scalar potential due to static charges distributed uniformly on
the cylindrical surface fitting the solenoid.

The distributions of $L$ as functions of the radial distance $r$ with $z=0$
and $z_{1}=-z_{2}$ are shown in Fig.\hspace{0.1cm}2. When the solenoid is
long, it is seen that the distribution of $L$ is almost uniform with respect
to $r$ when $r<r_{0}$. Thereby, for a cylindrical capacitor placed inside a
long rotating solenoid, the electromotance can vanish, as reported widely in
the literature. On the other hand, it is of essence to note that outside the
solenoid the distribution of $L$ departures from being uniform.
Consequently, it is predicted that a nonvanishing electromotance will be
induced on the stationary capacitor when it is placed outside the rotating
solenoid. Moreover, when the solenoid is not long, the distribution of $L$
is nonuniform even when $r<r_{0}$. Thus a nonvanishing electromotance can
also be induced on the capacitor when it is placed inside a short solenoid.
In the experiment of Kennard with an outside capacitor, a nonvanishing
electromotance was actually observed when the magnetization was reversed
[3]. However, this experiment cannot discriminate between the electromotance
due to the induction force with a time-varying current and the one due to
the magnetostatic force, which are given by the second and the first term on
the right-hand side of (21), respectively. However, a key difference between
them is that the contribution due to the electric induction force occurs
only at the instant of magnetization reversal and then disappears, while the
one due to the magnetostatic force remains after the reversal. Thus it is
possible to separate these two different kinds of electromotance with some
auxiliary devices such as switches.

The predicted electromotance tends to accumulate positive and negative
charges respectively on the two cylinders via the connecting wires. The
induced charges in turn generate the electric scalar potential. At each
point on the conducting plate, another scalar potential defined as the
electromotance with respect to a certain reference point can be given
unambiguously, as the electromotance due to the magnetostatic force and
given by (24) does not differentiate the paths connecting the two points.
Thereby, the distribution of charges can be determined by using the
condition that the difference between such a \textrm{magnetostatic potential}
and the electric scalar potential is fixed on the conducting plate. Thereby,
an accurate evaluation of the charge density can be given by solving a
relevant equation. Meanwhile, for a long solenoid the \textrm{magnetostatic
potential} can be expected to be uniform longitudinally. From Fig.\hspace{%
0.1cm}3, it is seen that for a solenoid with a length of 10 $r_{0}$, the
longitudinal distribution of the $L$ function can be fairly uniform near the
center of the solenoid with $|z|<r_{0}$. Then, by neglecting the fringing
effect and the longitudinal variation of the \textrm{magnetostatic potential}%
, the total charge induced on either cylinder of a short capacitor is then
given simply by $\pm Q$, where $Q=\mathcal{V}2\pi \epsilon _{0}l/\ln \left(
r_{b}/r_{a}\right) $, $l$ is the length of the cylindrical capacitor, and $%
\mathcal{V}$ is the electromotance between two suitable points on the
respective cylinders. The amount of the accumulated charges can be measured
if suitable \textrm{shielding} and switches are provided, as in Barnett's
experiment. This prediction of the nonvanishing electromotance and hence
induced charges seems not to be reported before and then provides a means to
test the validity of the modified force law.

For the case where both the solenoid and the capacitor are rotating, the
situation is a little complicated, as the wires connecting the rotating
capacitor to the electrometer are ordinarily stationary. Thus the path $C$
implicitly associated with (24) should be divided into segments, over each
of which its velocity is continuous as required in deriving (18). The
electromotance over a stationary segment is the same as given before. Then
we just consider a segment which as well as the solenoid is rotating at the
same rate $\omega $. Obviously, the electromotance as well as the velocity $%
\mathbf{v}_{em}$ of this case is the sum of those of the two cases where
either the segment or the solenoid is rotating. Thus for a long solenoid the
electromotance is expected to be still given by (12) when $r_{0}>r_{b}>r_{a}$%
, and by (25) and (26) when $r_{b}>r_{a}>r_{0}$, where $r_{a}$ and $r_{b}$
are associated with the endpoints of the segment. Meanwhile, the situation
can be quite different for the case with a short solenoid. Anyway, the
electromotance can be given by numerical calculation. For the corotating
case, the term $\mathbf{v}_{em}\mathbf{\cdot }d\mathbf{l}^{\prime }$ is
proportional to $\hat{r}^{\prime }\mathbf{\cdot }(\mathbf{r}^{\prime }-%
\mathbf{r})$, where $\hat{r}^{\prime }$ denotes a unit vector pointing from
the center axis to the various segments of the rotating solenoid at a given
instant. The electromotance is still given by (25) with the change that $L$
corresponds to the integral of $\hat{r}^{\prime }\mathbf{\cdot }(\mathbf{r}%
^{\prime }-\mathbf{r})/R$ and is then given by 
$$
L(r,z)=\frac{1}{2\pi (z_{2}-z_{1})}\int_{z_{1}}^{z_{2}}\int_{0}^{2\pi }\frac{%
r_{0}-r\cos \phi ^{\prime }}{\sqrt{(r_{0}\cos \phi ^{\prime
}-r)^{2}+r_{0}^{2}\sin ^{2}\phi ^{\prime }+(z^{\prime }-z)^{2}}}d\phi
^{\prime }dz^{\prime }.\eqno
(27) 
$$
The distributions of $L$ as functions of the radial distance $r$ with $z=0$
are shown in Fig.\hspace{0.1cm}4. It is seen that the numerical results
agree with the preceding estimate of the electromotance.

\vspace{1cm}

\noindent {\large \textbf{4. Conclusion}}\\[0.2cm]
Based on the modified Lorentz force law, which complies with Galilean
transformations and depends on the relative velocity between the effector
and the source particle, the electromotance over a conducting wire is
derived. The electromotance is composed of two parts which are due primarily
to the magnetostatic force and to the induction force, respectively. For a
closed wire the electromotance becomes the time derivative of the linked
magnetic flux, which is similar to the well-known formula related to
Faraday's law except that the translation, rotation, and deformation of the
loop have been taken into account explicitly. This time derivative can
vanish due to some symmetry, such as in Barnett's experiment with a circular
solenoid rotating about its center axis. Thus the electromotance reduces to
the contribution due to the magnetostatic force and thus depends on the
relative velocity between the effector and the matrix and on the positions
of the endpoints of the wire. For the case with a long solenoid, the \textrm{%
magnetostatic potential} is uniform inside the solenoid. Thereby, the
electromotance induced on a cylindrical capacitor placed inside a rotating
solenoid vanishes. This agrees with Barnett's experiment. However, for the
case with a short solenoid or for the case where the concentric capacitor is
placed outside the solenoid, the electromotance can be different from zero.
This prediction of nonvanishing electromotance then provides a means to test
the validity of the modified Lorentz force law.

\vspace{1.5cm}

\noindent {\large \textbf{References}}

\begin{itemize}
\item[{\lbrack 1]}]  S.J. Barnett, ``On electromagnetic induction and
relative motion,'' \textit{Phys. Rev}., vol. 35, pp. 323--336, Nov. 1912.

\item[{\lbrack 2]}]  E.H. Kennard, ``On unipoar induction: Another
experiment and its significance as evidence for the existence of the
aether,'' \textit{Philos. Mag}., ser. 6, vol. \textrm{33}, pp. 179--190,
1917.

\item[{\lbrack 3]}]  E.H. Kennard, ``Unipolar induction,'' \textit{Philos.
Mag}., ser. 6, vol. 23, pp. 937--941, \textrm{June} 1912.

\item[{\lbrack 4]}]  J. Djuri\'{c}, ``Spinning magnetic fields,'' \textit{J.
Appl. Phys}., vol. 46, pp. 679--688, Feb. 1975.

\item[{\lbrack 5]}]  D.F. Bartlett, J. Monroy, and J. Reeves, ``Spinning
magnets and Jehle's model of the electron,'' \textit{Phys. Rev. D}, vol. 16,
pp. 3459--3463, Dec. 1977.

\item[{\lbrack 6]}]  A.L. Kholmetskii, ``One century later: Remarks on the
Barnett experiment,'' \textit{Am. J. Phys}., vol. 71, pp. 558--561, June
2003.

\item[{\lbrack 7]}]  C.C. Su, \textit{Quantum Electromagnetics -- A
Local-Ether Wave Equation Unifying Quantum Mechanics, Electromagnetics, and
Gravitation} (2005),\newline
\texttt{http://qem.ee.nthu.edu.tw}

\item[{\lbrack 8]}]  C.C. Su, ``Connections between the Riemann and the
Lorentz force law,'' to appear in \textit{Found. Phys. Lett.,} vol. 19, pp.
187--194, Apr. 2006;\newline
\texttt{http://arXiv.org/physics/0506043}

\item[{\lbrack 9]}]  E.T. Whittaker, \textit{A History of the Theories of
Aether and Electricity} (Amer. Inst. Phys., New York, 1987), vol. 1, ch.%
\hspace{0.1cm}7.

\item[{\lbrack 10]}]  J.J. Thomson, {``On the electric and magnetic effects
produced by the motion of electrified bodies,'' \textit{Philos. Mag}., ser.
5, vol. 11, pp. 229--249, \textrm{Apr}. 1881.}

\item[{\lbrack 11]}]  J.D. Jackson, \textit{Classical Electrodynamics}
(Wiley, New York, 1999), ch.\hspace{0.1cm}5.
\end{itemize}

\newpage

\hspace{2.4cm}\includegraphics[bb=0 0 4in 3in]{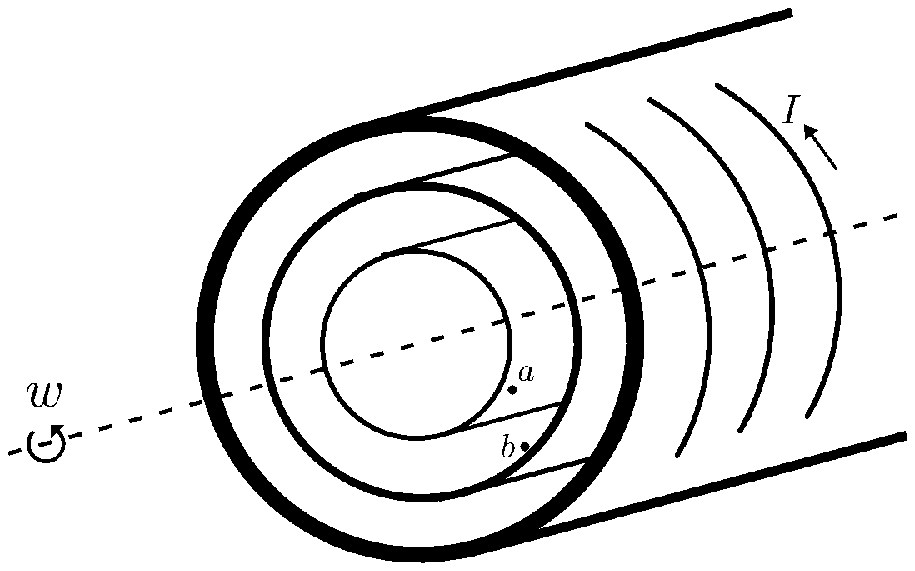}\vspace{0.3cm}

\hspace{1.1cm}%
\parbox{5in} {\baselineskip 0.1667in 
{\bf Fig.\hspace{0.1cm}1\hspace{0.2cm}} 
Schematic of Barnett's experiment with a solenoid carrying a current $I$ and a cylindrical capacitor. The inner and outer cylinders of the capacitor are coaxial with the solenoid and are connected to an electrometer with conducting wires and switches (not shown). The electromotance between point $a$ on the inner cylinder and point $b$ on the outer one is concerned.
}\vspace{1cm}

\hspace{0.8cm}\includegraphics[bb=0 0 6in 4in]{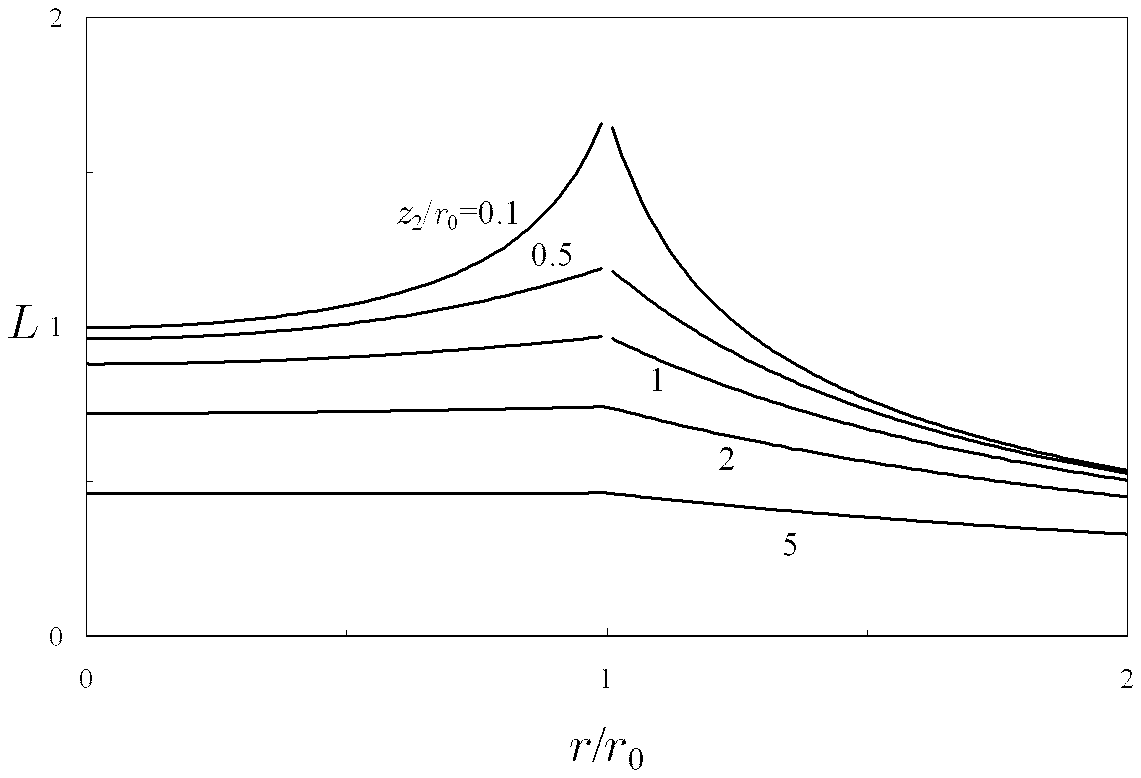}\vspace{0cm}

\hspace{1.1cm}%
\parbox{5in} {\baselineskip 0.1667in 
{\bf Fig.\hspace{0.1cm}2\hspace{0.2cm}} 
Radial distribution of the $L$ function at the center of the solenoid ($z=0$) with its length ($2z_{2}$) as a parameter. The solenoid is rotating and the capacitor is stationary.
}

\newpage

\hspace{0.8cm}\includegraphics[bb=0 0 6in 4in]{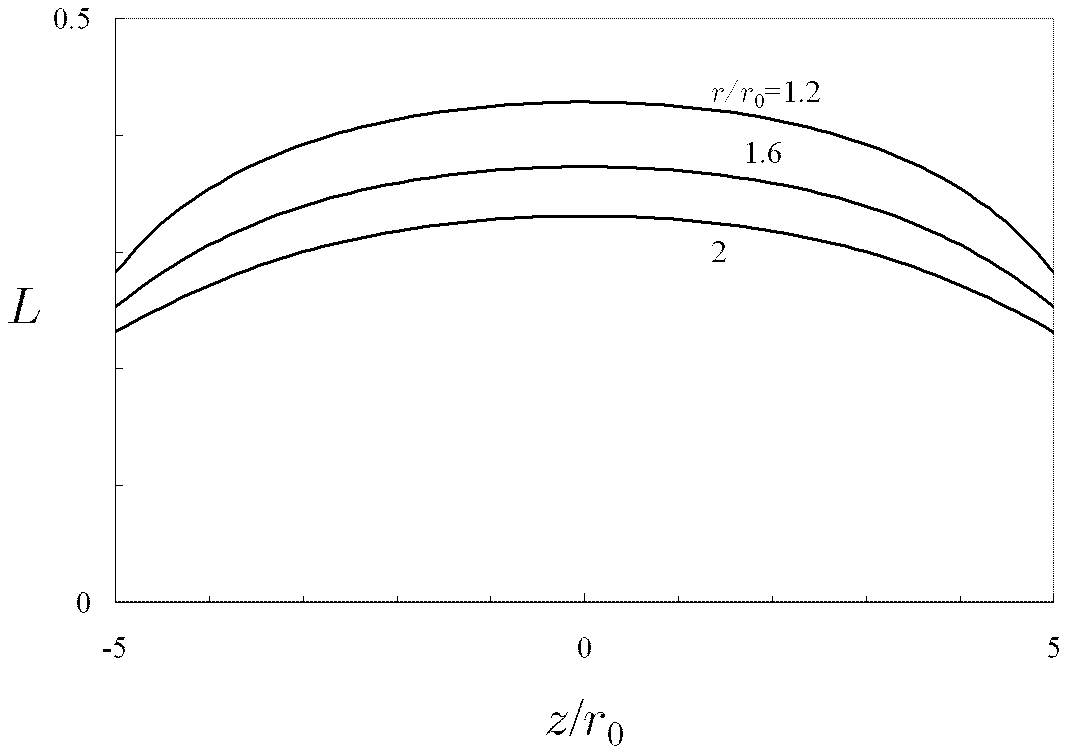}\vspace{0cm}

\hspace{1.1cm}%
\parbox{5in} {\baselineskip 0.1667in 
{\bf Fig.\hspace{0.1cm}3\hspace{0.2cm}} 
Longitudinal distribution of the $L$ function at $r=1.2$, 1.6, 2 $r_{0}$. The solenoid is centered at $z=0$ and of length 10 $r_{0}$.
}\vspace{1cm}

\hspace{0.8cm}\includegraphics[bb=0 0 6in 4in]{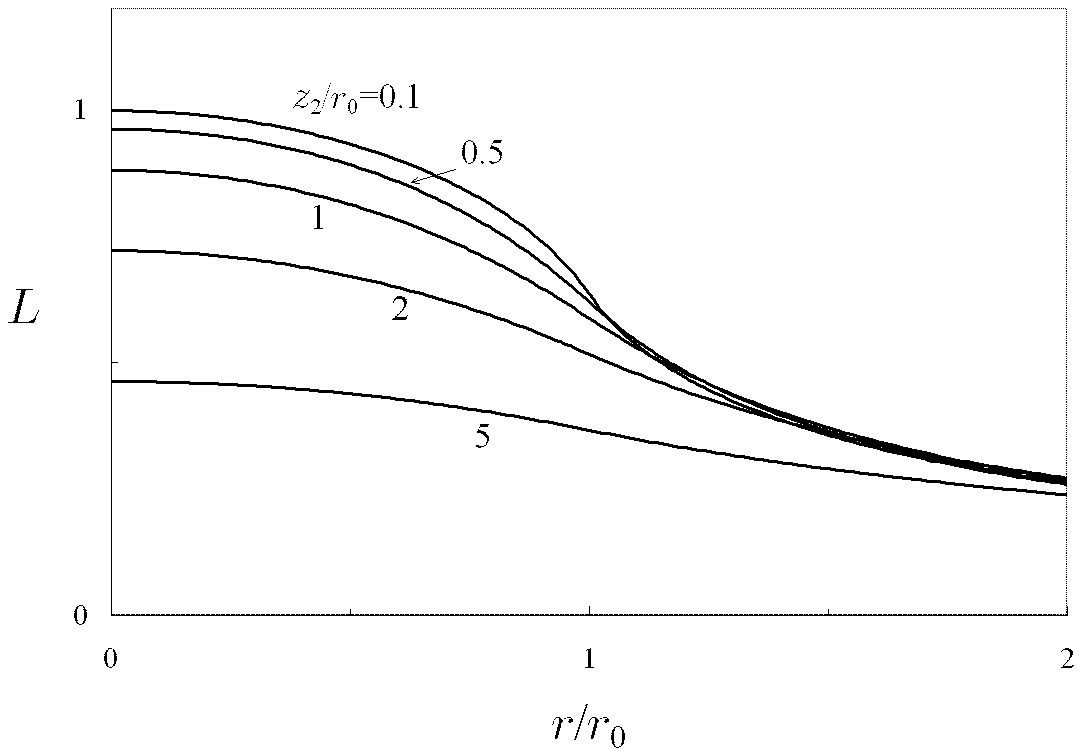}\vspace{0cm}

\hspace{1.1cm}%
\parbox{5in} {\baselineskip 0.1667in 
{\bf Fig.\hspace{0.1cm}4\hspace{0.2cm}} 
Radial distribution of the $L$ function at the center of the solenoid. Both the solenoid and the capacitor are rotating.
}

\end{document}